\documentclass[amsmath,amssymb,aps,10pt,prd,letterpaper,nofootinbib,balancelastpage,notitlepage,superscriptaddress,floatfix]{revtex4-2}
\usepackage{graphicx}	
\usepackage{ragged2e}	
\usepackage{bm}		    
\usepackage[sort&compress]{natbib}	 	
	\setcitestyle{square,numbers,comma}	
\usepackage[colorlinks=true,urlcolor=blue,linkcolor=red,citecolor=red]{hyperref}	
\newcommand{\tabref}[2][]{Tab{#1}.~\ref{tab:#2}}		
\newcommand{\figref}[2][]{Fig{#1}.~\ref{fig:#2}}		
\newcommand{\secref}[2][]{Sec{#1}.~\ref{sec:#2}}		
\newcommand{\appref}[2][x]{Appendi{#1}~\ref{app:#2}}	
\renewcommand{\eqref}[2][]{Eq{#1}.~(\ref{eq:#2})}		
\newcommand{\eqrefRange}[2]{Eqs.~(\ref{eq:#1})--(\ref{eq:#2})}		
\newcommand{\citeR}[2][]{Ref{#1}.~\cite{#2}}			
\newcommand{\A}{\bm A}
\newcommand{\B}{\bm B}
\newcommand{\dd}[2]{\frac{\partial#1}{\partial#2}}
\newcommand{\DM}{\mathrm{DM}}
\newcommand{\E}{\bm E}
\newcommand{\eV}[1][]{\,\mathrm{#1eV}}
\newcommand{\ga}{g_{a\gamma}}
\newcommand{\Hz}[1][]{\,\mathrm{#1Hz}}
\let\Im\relax\DeclareMathOperator{\Im}{\text{Im}}
\newcommand{\jeff}[1][]{J_{\mathrm{eff}#1}}
\newcommand{\Jeff}[1][]{\bm J_{\mathrm{eff}#1}}
\renewcommand{\k}{\bm k}
\newcommand{\km}{\,\mathrm{km}}
\newcommand{\LLint}[1][]{\mathcal L_{\mathrm{int}#1}}
\newcommand{\nl}{\nonumber \\ & \quad }
\renewcommand{\O}{\mathcal O}
\newcommand{\phihat}{\bm{\hat{\varphi}}}
\let\Re\relax\DeclareMathOperator{\Re}{\text{Re}}
\newcommand{\rhat}{\bm{\hat{r}}}
\newcommand{\rhoeff}[1][]{\rho_{\mathrm{eff}#1}}
\newcommand{\thetahat}{\bm{\hat{\theta}}}
\newcommand{\Tcoh}{T_\mathrm{coh}}
\newcommand{\x}{\bm x}

\begin{document}

\title{Earth as a transducer for ultralight bosonic dark-matter detection}
\date{\today}
\author{Saarik Kalia}
\email{skalia@ifae.es}
\affiliation{Institut de F\'isica d’Altes Energies (IFAE), The Barcelona Institute of Science and Technology, Campus UAB, 08193 Bellaterra (Barcelona), Spain}
\author{Ibrahim A. Sulai}
\email{ibrahim.sulai@bucknell.edu}
\affiliation{Department of Physics \& Astronomy, Bucknell University, Lewisburg, Pennsylvania 17837, USA}

\begin{abstract}
Ultralight bosonic dark matter (UBDM) that couples to electromagnetism can generate an oscillating magnetic-field signal at the Earth's surface.  This is referred to as the ``Earth transducer" effect, as the Earth converts UBDM into a detectable magnetic field.  Similar DM-induced fields in laboratory experiments typically scale with the size $L$ of the experiment.  Because the Earth transducer signal instead scales with the large radius of the Earth, $R$, it is one of the most powerful direct probes of UBDM with masses $m_\DM\lesssim1/R\sim3\times10^{-14}\eV$.  It has many other favorable properties, such as high spatial and temporal coherence and robustness to atmospheric modeling.  In this review, we derive the Earth transducer effect and its properties for multiple UBDM models, and discuss current and future prospects to detect it.
\end{abstract}

\maketitle

\section{Introduction}
\label{sec:introduction}

An abundance of astrophysical and cosmological observations indicate the presence of a poorly understood component of the universe that we call dark matter (DM).  Its existence is inferred from its gravitational impact on a variety of phenomena, including the motion of stars/galaxies~\cite{Zwicky:1933gu,Rubin:1970zza}, the merging of clusters~\cite{Clowe_2006}, and the expansion of the universe~\cite{Planck:2018vyg}.  While these observations indicate its approximate abundance and velocity~\cite{Riccardo_Catena_2010,Evans:2018bqy,S_ding_2025}, very little else is known about DM.  Some of its properties of interest include the fundamental mass and spin of its constituents and its possible non-gravitational interactions.

Of the myriad models which have been proposed for DM, one class which has garnered significant interest recently is ultralight bosonic dark matter (UBDM)~\cite{Arias:2012az,kimball2022search,Antypas:2022asj}.  This class consists of models where DM is composed of constituent particles with masses $m_\DM\lesssim10\eV$.%
\footnote{We work in natural units $\hbar=c=\varepsilon_0=1$.}
In this regime, the DM behaves as a classical field~\cite{lin2018self,Centers:2019dyn,cheong2024}.  Popular UBDM candidates include dark-photon dark matter (DPDM)~\cite{Holdom:1986ag,cvetivc1996implications,Nelson:2011sf,Graham:2015rva}, axionlike-particle dark matter (henceforth, simply axion DM)~\cite{Preskill:1982cy,Abbott:1982af,Dine:1982ah,Svrcek:2006yi,Arvanitaki:2009fg}, and millicharged dark matter (mDM)~\cite{Holdom:1986ag,Wen_1985,Hall_2010,Alonso_lvarez_2019,Jaeckel_2021}.

All of these candidates may couple to electromagnetism (EM)~\cite{Holdom:1986ag,Sikivie:1983ip}, and a variety of laboratory experiments have been conducted or proposed to search for such interactions~\cite{Jaeckel_2009,ehret2010new,Redondo:2010dp,Horns:2012jf,Betz_2013,Chaudhuri:2014dla,Caldwell:2016dcw,Anastassopoulos:2017ftl,Baryakhtar:2018doz,armengaud2019physics,Lawson:2019brd,Berlin_2020_dd,Berlin_2020_srf,gramolin2021search,Andrianavalomahefa:2020ucg,Salemi:2021gck,Berlin_2022,Chiles_2022,Adair_2022,Jiang_2024,Berlin_2024,Bai_2025,Carosi_2025,Kalia:2025afc}.  The most widely explored strategy is the use of resonant cavities~\cite{Sikivie:1983ip,Adair_2022,Bai_2025,Carosi_2025}.  A key feature of such experiments is that the size $L$ of the cavity should match the Compton wavelength $\lambda_\DM\equiv2\pi/m_\DM$ of the DM, which becomes increasing difficult below $m_\DM\sim\mu\mathrm{eV}$.  At lower masses, the use of resonant LC circuits can be employed~\cite{Chaudhuri:2014dla,Salemi:2021gck}.  The signal in these systems scales with the size $L$ of the experimental shield, when $\lambda_\DM\gg L$.

In general, the presence of a conducting boundary around a laboratory experiment imposes a limitation to detecting these models of UBDM in the extremely low mass regime.  A natural question is whether any advantage can be gained by removing the experimental shield entirely.  In \citeR{dpdm_theory}, it was shown that in this case, the Earth sets the conducting boundary of the system.  This is because both the ground and ionosphere act as good conductors at extremely low frequencies.  The dominant signal in this system is then an oscillating magnetic field at the Earth's surface, whose magnitude scales with the radius $R$ of the Earth.  This effect is known as the ``Earth transducer" effect, as the Earth can convert UBDM into a detectable magnetic-field signal.

The Earth transducer effect provides a new method for direct detection of UBDM coupled to EM, namely via correlated measurements from a global array of unshielded magnetometers.  The UBDM signal would appear as a sharp peak in the spectrum of each magnetometer, whose amplitude exhibits a prescribed dependence on the magnetometer's geographic location.  These unique properties make the signal easily distinguishable from noise.  There are a number of completed and ongoing initiatives to utilize existing datasets and undertake dedicated searches to look for this effect.  \citeR[s]{dpdm_search,axion,supermag_1sec} performed searches in publicly available geomagnetic-field datasets maintained by the SuperMAG collaboration~\cite{SuperMAGwebsite,Gjerloev:2009wsd,Gjerloev:2012sdg}, while \citeR[s]{eskdalemuir_letter,eskdalemuir_analysis,eskdalemuir_dpdm} searched data collected from the Eskdalemuir observatory~\cite{eskdalemuir}.  In addition, the SNIPE Hunt collaboration~\cite{snipe_hunt,snipe_hunt_2026}, Munich group~\cite{munich}, and GPEX collaboration~\cite{gpex} have taken correlated measurements with unshielded magnetometers and analyzed their datasets to search for the Earth transducer effect.  In many regions of parameter space, these searches are the leading direct constraints on certain UBDM models (see \figref[s]{constraints} and \ref{fig:constraints_mdm}).

In this review, we summarize the current progress and outlook for probing UBDM parameter space with the Earth transducer effect.  In \secref{theory}, we begin by describing the UBDM models of interest and deriving the Earth transducer effect and its properties for these models.  We also comment on methods to extend the signal modeling to higher frequencies for which $\lambda_\DM\lesssim R$.  In \secref{experiment}, we review the searches for this effect in existing datasets and dedicated searches.  The results from these searches are summarized in \figref[s]{constraints} and \ref{fig:constraints_mdm}.  In \secref{conclusion}, we discuss the current state of these detection efforts and conclude.  In \appref{VSH}, we review the relevant properties of vector spherical harmonics (VSH).

\section{Theory}
\label{sec:theory}

In this section, we review the theoretical framework for the Earth transducer effect.  This framework was first developed for DPDM in \citeR{dpdm_theory}, and then extended to axion DM and mDM in \citeR[s]{axion,millicharged}, respectively.  We begin by reviewing properties of EM-coupled UBDM models and deriving how they manifest in experimental contexts.  All of these models can be described in a common framework, namely as an ``effective current" $\jeff^\mu$.  In this language, standard techniques for analyzing EM problems can be applied to understand the effects of UBDM.  In particular, we will see that the effects of UBDM generically scale with the size $L$ of the experimental shield, when $\lambda_\DM\gg L$.

We then proceed to derive the Earth transducer effect.  We briefly discuss the conductive properties of the Earth's interior and atmosphere, in order to understand the EM boundary conditions of the system.  We then compute the Earth transducer effect for each UBDM candidate in a simple spherical model of the Earth's atmosphere.  Importantly, we show that the effect scales with the radius $R$ of the Earth, rather than the much smaller height $h$ of the lower atmosphere.  We also discuss the coherence properties of the signal which make it distinguishable from noise.  Although we compute the effect for a simplified model, we argue that for low masses, this result can be made robust to the environmental details of the Earth's atmosphere.

Finally, we review methods to extend the Earth transducer effect to higher masses, where this robustness breaks down.  \citeR{curl} proposed to perform a differential measurement of $\nabla\times\bm B$, in order to robustly probe the local effective current, indepedent of global boundary conditions.  Meanwhile, \citeR[s]{eskdalemuir_letter,eskdalemuir_theory,eskdalemuir_dpdm} computed the Earth transducer effect in a more refined empirical model of the Earth's atmospheric conductivity profile.

\subsection{UBDM models}
\label{sec:ubdm}

A number of intriguing models have been proposed for the fundamental nature of DM.  In this review, we will focus on so-called UBDM models~\cite{Arias:2012az,kimball2022search,Antypas:2022asj}, where the constituent particles are sufficiently light that the DM behaves like a classical field.  Astrophysical observations indicate that the local DM in the vicinity of the Solar System has energy density $\rho_\DM\sim m_\DM n_\DM\sim0.3\eV[G]/\mathrm{cm}^3$ and velocity $v_\DM\sim10^{-3}$~\cite{Riccardo_Catena_2010,Evans:2018bqy,S_ding_2025}.  Therefore, if $m_\DM\lesssim10\eV$, there exist $N_\DM\gtrsim1$ DM particles within each de Broglie volume $\lambda_\mathrm{dB}^3\equiv(2\pi/m_\DM v_\DM)^3$.  This implies two important properties for this DM mass range.  The first is that DM candidates in this mass range must be bosonic (as fermions are restricted to occupation numbers $N_\DM<1$).  The second is that quantum effects become unimportant and the DM behaves classically.

More specifically, the DM can be described by a classical field, such as a scalar (spin 0) field $\chi(\x,t)$, or a vector (spin 1) field $X_\mu(\x,t)$.%
\footnote{Throughout this review, we write $x^\mu$ for four-vectors and $\bm x$ for three vectors.}
To understand its kinematics, let us first focus on the case of a real scalar field $\chi$.  In the absence of interactions, this field satisfies the Klein-Gordon equation
\begin{equation}\label{eq:KG}
    \square\chi+m_\DM^2\chi=0.
\end{equation}
As this is a type of wave equation, UBDM is sometimes also referred to as ``wave DM".  Generically, \eqref{KG} is solved by a sum of plane waves
\begin{equation}\label{eq:plane}
    \chi(\x,t)=\Re\left[\sum_n\chi_{0,n}e^{i(\k_n\cdot\x-\omega_nt)}\right],
\end{equation}
where $\omega_n^2=|\k_n|^2+m_\DM^2$, and $\chi_{0,n}$ are various complex amplitudes.  As $|\k_n|/m_\DM\sim v_\DM\ll1$, then \eqref{plane} simplifies to the form
\begin{equation}\label{eq:plane_nr}
    \chi(\x,t)\approx\chi_0\cos(m_\DM t+\alpha_0),
\end{equation}
for a single real amplitude $\chi_0$ and phase $\alpha_0$.  In other words, we expect the DM to be spatially homogeneous and oscillating in time with frequency $f_\DM\equiv m_\DM/2\pi$.  Note that the approximation in \eqref{plane_nr} only applies for distances within a de Broglie wavelength, $|\x|\lesssim\lambda_\mathrm{dB}$, and times within a coherence time, $t\lesssim\Tcoh\equiv 1/(f_\DM v_\DM^2)$.  Over larger length- or timescales, $\chi_0$ and $\alpha_0$ will generically vary.  The typical size of $\chi_0$ is fixed by the local DM energy density%
\footnote{For a vector field $X_\mu$, the corresponding relation is $\langle|X_{i,0}|^2\rangle=2\rho_\DM/3m_\DM^2$, for each spatial component $i=1,2,3$.}
\begin{equation}
    \langle\chi_0^2\rangle=\frac{2\rho_\DM}{m_\DM^2},
\end{equation}
where $\langle\cdot\rangle$ represents an average over several de Broglie wavelengths or coherence times.

In this review, we will consider UBDM models which couple to EM.  All such models can be treated in a common framework.  For any interaction $\LLint$ between a new field and the Standard Model (SM) photon $A_\mu$, we may define an ``effective current"
\begin{equation}\label{eq:jeff}
    \jeff^\mu\equiv\partial_\nu\dd\LLint{(\partial_\nu A_\mu)}-\dd\LLint{A_\mu}.
\end{equation}
Under a gauge transformation $A_\mu\rightarrow A_\mu+\partial_\mu\alpha$, the interaction term transforms as
\begin{equation}
    \LLint\rightarrow\LLint-\jeff^\mu\partial_\mu\alpha.
\end{equation}
Gauge invariance then implies that the effective current must be conserved, i.e. $\partial_\mu\jeff^\mu=0$.  This effective current behaves similarly to a SM current.  In particular, it modifies Maxwell's equations as
\begin{align}\label{eq:gauss}
    \nabla\cdot\E&=\rhoeff,\\
    \nabla\cdot\B&=0,\\
    \nabla\times\E+\partial_t\B&=0,\\
    \nabla\times\B-\partial_t\E&=\Jeff,
\label{eq:ampere}\end{align}
where $\jeff^\mu=(\rhoeff,\Jeff)$.  In the limit that the DM is very weakly coupled to the SM, backreaction can be neglected and the DM can be treated as a background field.  Furthermore, if background EM fields are present, then all instances of $A_\mu$ appearing in $\jeff^\mu$ can take their background values.  Assuming these EM fields are static, the effective current will inherit the oscillatory time dependence of the DM.  Then to leading order, $\jeff^\mu$ will behave as a background AC current density, and standard EM techniques can be applied to understand its observable effects.

In this review, we will consider three UBDM models.  The first is a massive vector field $A'_\mu$, called a (kinetically mixed) dark photon.  This candidate couples to EM via the interaction%
\footnote{The DPDM interaction is often expressed instead as a ``kinetic mixing", $\LLint[,A']=\frac\varepsilon2F'_{\mu\nu}F^{\mu\nu}$.  To leading order, this is equivalent to \eqref{DPDM_coupling} via a field redefinition $A_\mu\rightarrow A_\mu,A'_\mu\rightarrow A'_\mu+\varepsilon A_\mu$.}
\begin{equation}\label{eq:DPDM_coupling}
    \LLint[,A']=\varepsilon m_{A'}^2A_\mu A'^\mu,
\end{equation}
where $m_{A'}$ is the mass of the DPDM field,%
\footnote{In contexts which are relevant to a particular model, we write the mass as $m_{A'}$, $m_a$, or $m_\phi$.  In contexts which apply to all UBDM  models, we write $m_\DM$.}
and $\varepsilon<1$ parametrizes its coupling to EM.  The equation of motion for $A'_\mu$ is the Proca equation,
\begin{equation}\label{eq:Proca}
    \partial_\mu F'^{\mu\nu}+m^2A'^\nu=0,
\end{equation}
where $F'_{\mu\nu}=\partial_\mu A'_\nu-\partial_\nu A'_\mu$ is the DPDM field strength.  Taking the derivative of \eqref{Proca} shows that $\partial_\mu A'^\mu=0$.  This implies that $A'_0\sim v_\DM\cdot\A'$, so the temporal components of $A'_\mu$ are suppressed.  The spatial components take an analogous form to \eqref{plane_nr},
\begin{equation}\label{eq:plane_DPDM}
    A'_i(\x,t)=A'_{i,0}\cos(m_{A'}t+\alpha_{i,0}),
\end{equation}
where $A'_{i,0}$ and $\alpha_{i,0}$ are amplitudes and phases for each component $i=1,2,3$.  Importantly, the phases $\alpha_{i,0}$ need not be the same.  This allows $\A'(t)$ to exhibit an elliptical polarization (in which case $|\A'(t)|\neq0$ throughout its entire oscillation).  From \eqref[s]{jeff} and (\ref{eq:DPDM_coupling}), we see that the effective current for DPDM is given by
\begin{equation}\label{eq:jeff_DPDM}
    \jeff[,A']^\mu=-\varepsilon m_{A'}^2A'^\mu.
\end{equation}
As with $A'_\mu$, the temporal component of \eqref{jeff_DPDM} is suppressed, i.e. there is no effective charge $\rhoeff$, and the spatial components are uniform in space.  \eqref{jeff_DPDM} also makes it clear that the observable effects of DPDM should scale with $m_{A'}$.  This makes detection of DPDM particularly difficult in the low-mass regime.

Our second candidate is a real scalar field $a$, known as an axionlike particle (or simply axion).  As a real scalar, it satisfies \eqref{KG} and takes the form in \eqref{plane_nr}.  This candidate couples to EM via the interaction
\begin{equation}
    \LLint[,a]=\frac14\ga aF_{\mu\nu}\tilde F^{\mu\nu},
\end{equation}
where $\ga$ is the axion-photon coupling, and $\tilde F_{\mu\nu}=\frac12\epsilon^{\mu\nu\rho\sigma}F_{\rho\sigma}$.  From \eqref{jeff}, the effective current corresponding to this interaction is
\begin{equation}
    \jeff[,a]^\mu=-\frac12\ga\epsilon^{\mu\nu\rho\sigma}\partial_\nu aF_{\rho\sigma},
\end{equation}
where $\epsilon^{\mu\nu\rho\sigma}$ is the Levi-Civita symbol, or equivalently,
\begin{align}
    \rhoeff[,a]&=\ga\nabla a\cdot\B,\\
    \Jeff[,a]&=-\ga[(\partial_ta)\B+\nabla a\times\E].
\label{eq:jeff_axion}\end{align}
Unlike the case of DPDM, the axion effective current depends on the EM field.  In order to enhance the observable effects of axion DM, it is therefore beneficial to introduce background EM fields.  Because $\nabla a\sim v_\DM\cdot\partial_ta$, we see that the largest effect will arise from the first term in \eqref{jeff_axion}, and so we should introduce a background magnetic field $\B_0$.  In the case of the Earth transducer effect, this role will be played be the Earth's DC magnetic field.  In contrast to $\Jeff[,A']$, which is uniform in space, $\Jeff[,a]$ will follow the dipolar shape of the geomagnetic field.

Our final model of interest is a complex scalar field $\phi$, called a millicharged particle.  This field is charged under EM and so couples through its kinetic term
\begin{equation}
    \LLint[,\phi]=D_\mu\phi(D^\mu\phi)^*,
\end{equation}
where $D_\mu=\partial_\mu+ie_m A_\mu$, and $e_m$ is the electric charge of the mDM.  The corresponding effective current is given by
\begin{equation}\label{eq:jeff_mdm_full}
    \jeff[,\phi]^\mu=ie_m(\phi^*\partial^\mu\phi-\phi\partial^\mu\phi^*)-2e_m^2|\phi|^2A^\mu.
\end{equation}
As the effective current depends explicitly on $A_\mu$, it is important to take care with the choice of gauge.  Physically, we expect $\rhoeff=0$, so that the mDM does not generate large-scale electric fields throughout the galaxy.  This implies the condition
\begin{equation}\label{eq:no_charge}
    \Im[\phi\partial_t\phi^*]=e_m|\phi|^2A_0.
\end{equation}
Fixing temporal Coulomb gauge, $A_0=0$ and $\nabla\cdot\A=0$,%
\footnote{In all contexts in this review, there will be no net SM charge density, $\rho_\mathrm{SM}=0$, so it is consistent to impose both of these conditions.}
we then see that \eqref[s]{KG} and (\ref{eq:no_charge}) imply the form%
\footnote{\label{ftnt:deflection}%
If $e_m$ is sufficiently large, the mDM can be deflected by the geomagnetic field.  This effect can significantly alter the background distribution of the mDM.  The distribution in \eqref{plane_mdm} [and the resulting signal in \eqref{mdm_signal}] remain valid so long as $e_m^2|\B_0|^2R^4/4\ll1$~\cite{snipe_hunt_mdm}.}
\begin{equation}\label{eq:plane_mdm}
    \phi(\x,t)=\phi_0\cos(m_\phi t+\alpha_0),
\end{equation}
in the non-relativistic (NR) limit $v_\DM\ll1$.  In principle, \eqref{no_charge} can be satisfied even for a complex amplitude $\phi_0$, but the phase of $\phi_0$ can always be removed by a gauge transformation, so we will take it to be real.  With the form in \eqref{plane_mdm}, the spatial components of the effective current become
\begin{equation}\label{eq:jeff_mdm}
    \Jeff[,\phi]=-e_m^2\phi_0^2[1+\cos(2m_\phi t+2\alpha_0)]\A
\end{equation}
in the NR limit.  This effective current contains both DC and AC contributions.  In this review, we will focus on the effect of the AC part.  Note that in the mDM case, the AC part of the effective current oscillates at frequency $2m_\phi$, as will any response fields.  This is because the relevant contribution to $\jeff[,\phi]^\mu$ in \eqref{jeff_mdm_full} depends on $|\phi|^2$.  Again, to generate an observable effect, we require a background EM field, which will be the geomagnetic field in the case of the Earth transducer effect.

Now that we have unified all of our UBDM models into a common effective current framework, let us discuss the possible effects which could be observed in an experiment.  All of our models have $\rhoeff=0$, so the most relevant of Maxwell's equations will be the Amp\`ere-Maxwell law, given by \eqref{ampere}.  In general, the AC effective current will source AC electric fields $\E_\DM$ and/or magnetic fields $\B_\DM$ inside an experimental apparatus.  In this review, we will primarily focus on the extremely low mass regime, where the Compton wavelength of the DM exceeds the size of the experimental apparatus.  More specifically, suppose the experiment is surrounded by some conducting boundary of characteristic length $L$.  In the regime $\lambda_\DM\gg L$, there is a simple argument that $\E_\DM$ will generally be suppressed relative to $\B_\DM$ (see \citeR[s]{dpdm_theory,curl} for more detailed arguments in the Earth case).  Standard EM boundary conditions mandate that (the parallel component of) $\E_\DM$ must vanish at the conducting boundary.  As $\E_\DM$ has frequency $m_\DM$, it should grow on length scales $\lambda_\DM$.  The electric field at the center of the cavity can then be at most $\O(L^2/\lambda_\DM^2)$ [whereas we will see that $\B_\DM$ scales as $L^1$].  Neglecting the electric field contribution, the Amp\`ere-Maxwell law then becomes
\begin{equation}\label{eq:ampere_noE}
    \nabla\times\B_\DM=\Jeff.
\end{equation}
Parametrically, this implies $\B_\DM\sim\Jeff\cdot L$.  In summary, in the limit $m_\DM L\ll1$, we expect the dominant signal to be an AC magnetic field whose magnitude scales with $L$.  We therefore benefit from utilizing as large of an apparatus as possible.

\subsection{Earth transducer effect}
\label{sec:transducer}

We have seen that the presence of a small conducting shield can suppress the signal from UBDM.  A natural question is: how large can we make our experimental shield, or can we even remove the shield entirely?  Inevitably, any experimental setup will have some conducting boundary.  If the experiment does not contain a deliberate shield, the room inside which the experiment is conducted may act as a shield~\cite{Jiang_2024}.  Even if the experiment is conducted outside a laboratory, the Earth itself will set a conducting boundary.  This is because the Earth's interior and the upper layers of the Earth's atmosphere act as good conductors at the low frequencies, $7\times10^{-4}\Hz\lesssim f_\DM\lesssim7\Hz$, corresponding to our mass range of interest, $3\times10^{-18}\eV\lesssim m_\DM\lesssim3\times10^{-14}\eV$.

The full details of the conductivity profile in/around the Earth are quite complex (see \citeR{dpdm_theory} for more detailed discussion).  Here we review the relevant details for our scenario.  A conductive layer of the Earth's interior or atmosphere acts as an efficient shield if its conductivity is higher than the frequency of the DM response, $\sigma\gg m_\DM$, and the layer is thicker than its skin depth (damping length), $d\gg\delta\sim\sqrt{2/\sigma m_\DM}$.  On the other hand, if a layer satisfies $\sigma\ll m_\DM$ or $d\ll\delta$, it contributes negligible damping to the DM response, and so can be treated as vacuum.  While the crust of the Earth is not sufficiently thick to constitute an effective shield, the inner layers (mantle and core) of the Earth are certainly thick enough to function as a shield for our entire mass range of interest~\cite{Price_1970,Hutton_1976}.  On the other hand, the lower $\sim100\km$ of the Earth's atmosphere has very low conductivity~\cite{Handbook20.1}, and so can be treated as vacuum.  The situation becomes more complicated for the ionosphere -- an upper layer of the Earth's atmosphere which is highly ionized by ultraviolet radiation -- due to the anisotropy and temporal modulation of its conductivity~\cite{Takeda:1985hcf,GM118}.  For masses $m_\DM\gg10^{-16}\eV$ ($f_\DM\gg0.02\Hz$), the ionosphere is certainly thick enough to constitute an effective shield, in which case, it sets a relatively spherical outer conducting boundary for the system.  For lower masses, the thickness condition is satisfied for certain directional conductivities but not others, and so it becomes unclear if the ionosphere can effectively damp the DM response.  Outside the Earth's atmosphere, the interplanetary medium acts as a collisionless plasma with plasma frequency $\omega_p\sim10^{-10}\eV\gg m_\DM$~\cite{kallenrode2004space}. If the ionosphere does not set the outer boundary of the system, then the interplanetary medium will.  It is worth noting that the magnetopause, which dilineates the boundary between the Earth's atmosphere and the interplanetary medium, is highly aspherical~\cite{ShueSong,SibeckLin}, which, in principle, can lead to a complicated geometry for the system.

\begin{figure*}[t]
\includegraphics[width=0.4\textwidth]{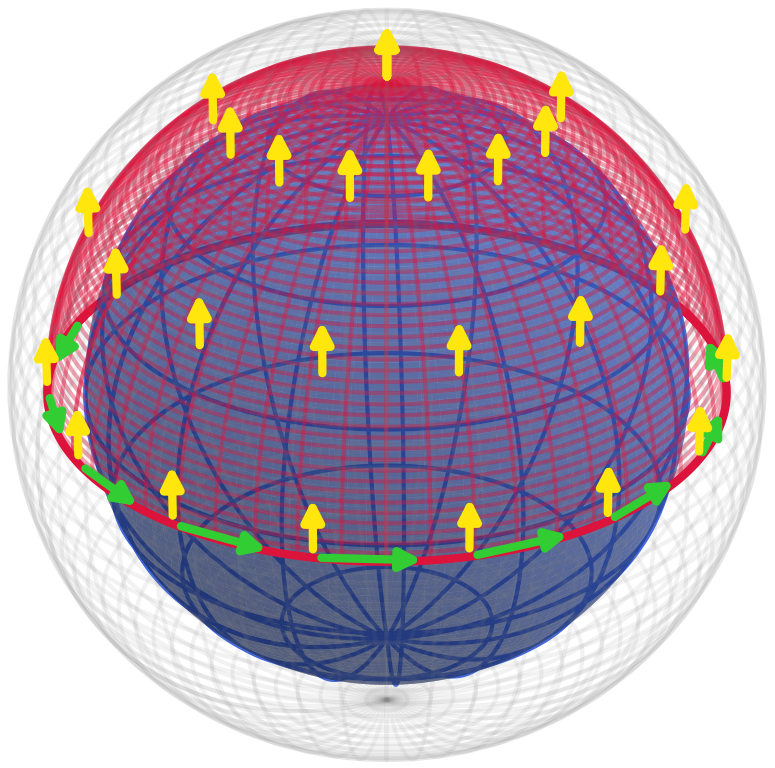}\hspace{0.07\textwidth}
\includegraphics[width=0.4\textwidth]{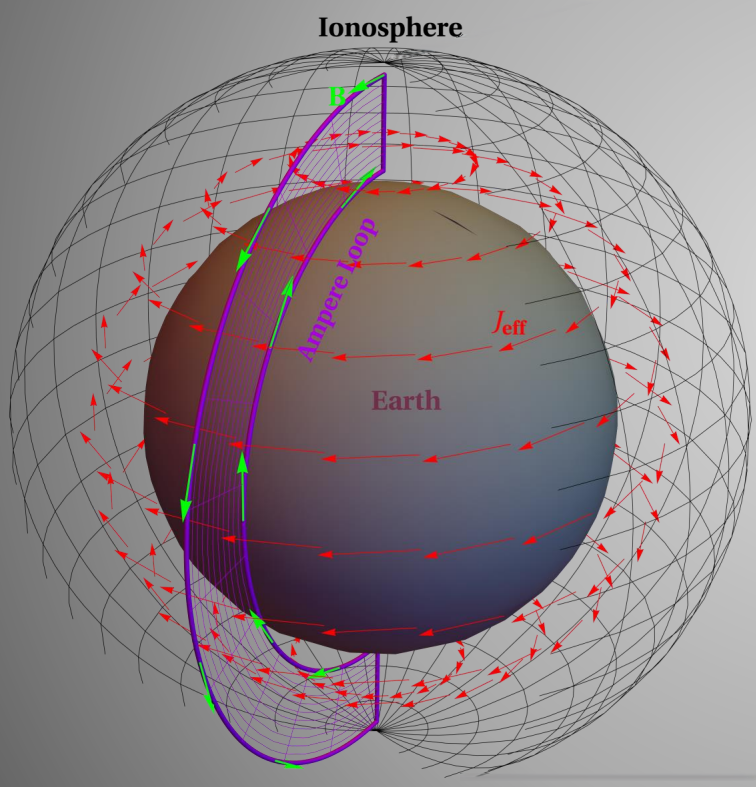}
\caption{\label{fig:ampere}%
    Sketches of idealized spherical Earth-ionosphere cavity and Amp\`ere's law arguments.  \emph{Left (from \citeR{dpdm_theory})}: The Earth and ionosphere are represented by the blue and grey spheres, respectively.  In red, we show a Gaussian surface covering the northern hemisphere.  The yellow arrows represent the DPDM effective current $\Jeff[,A']$, while the green arrows represent the magnetic-field response $\B_{A'}$ along the equator.  The Amp\`ere's law argument presented in \secref{transducer} implies $\B_{A'}\sim\Jeff[,A']\cdot R$.  \emph{Right (from \citeR{millicharged})}: Similar argument to left panel, but for mDM current $\Jeff[,\phi]$ soured by final term of \eqref{IGRF_A}.  Now the Amp\`erian loop is shown in purple and $\Jeff[,\phi]$ in red.  This argument demonstrates that an effective current of this form generates a suppressed response $\B_\phi\sim\Jeff[,\phi]\cdot h$.  The dominant response instead comes from the other terms in \eqref{IGRF_A}.}
\end{figure*}

To begin, let us compute the DM response for a simplified geometry of the Earth-ionosphere cavity, and then we will comment on how this result applies in the more realistic case.  We take the Earth to be a perfectly conducting sphere of radius $R=6400\km$, and the ionosphere to be a concentric perfectly conducting shell with inner radius $R+h$, so that the lower atmosphere is a vacuum region of height $h\sim100\km\ll R$.  We will solve \eqref{ampere_noE} in this geometry (for the appropriate $\Jeff$).  An important result will be that $\B_\DM\propto R$, as opposed to $\B_\DM\propto h$.  Before considering the full computation, let us review a simple geometric argument from \citeR{dpdm_theory}, shown in the left panel of \figref{ampere}, which demonstrates this fact in the case of DPDM.  Suppose that the DPDM is polarized along the Earth's rotational axis, so that $\Jeff[,A']$ also points in this direction.  Consider a Gaussian surface situated just above the Earth's surface, which covers the northern hemisphere.  The boundary of this surface is an Amp\`erian loop which wraps around the equator.  Amp\`ere's law, in integral form, states
\begin{equation}
    \Jeff\cdot R^2\sim\int \Jeff\cdot d\bm S=\oint \B_\DM\cdot d\bm l\sim\B_\DM\cdot R.
\end{equation}
It is then clear that $\B_\DM\sim\Jeff\cdot R$.  Note that this estimate does not depend on $h$.  In particular, the argument holds even in the limit $h\rightarrow0$.  This is important not only to show that the Earth transducer effect is enhanced by the large radius of the Earth, but also because the result is not sensitive to the exact value of $h$ (which depends on detailed atmospheric modeling).  In fact, the argument still applies even if the ionosphere is deformed away from a sphere (so long as $\E_\DM$ can still be ignored in the Amp\`ere-Maxwell law).  This demonstrates how the predictions of the Earth transducer effect can be made robust to environmental details of the Earth's atmosphere.

Now let us state the full results for each UBDM model.  In the DPDM case, solving \eqref{ampere_noE} in the spherical Earth-ionosphere cavity yields the result
\begin{equation}\label{eq:DPDM_signal}
    \bm B_{A'}(\Omega,t)=\sqrt{\frac{4\pi}{3}}\cdot\frac{m_{A'}R}{2-(m_{A'}R)^2}\cdot\mathrm{Re}\left[\sum_{m=-1}^1\varepsilon m_{A'}\tilde A_m^{'} \cdot\bm\Phi_{1m}(\Omega)\cdot e^{-i(m_{A'}-2 \pi f_dm)t}\right].
\end{equation}
Here $\bm\Phi_{\ell m}$ are VSH (see \appref{VSH}), and $\Omega=(\theta,\varphi)$ denotes geographic coordinates on the Earth's surface.%
\footnote{\label{ftnt:coords}%
Note that while $\varphi$ coincides with the usual definition of longitude, $\theta$ is not the usual latitude.  The geographic north pole (latitude $+90^\circ$) corresponds to $\theta=0$, while the geographic south pole (latitude $-90^\circ$) corresponds to $\theta=\pi$.}
In the NR limit, only the $\ell=1$ VSH are relevant for the DPDM signal.  The complex amplitudes $\tilde A'_m$ are related to the real Cartesian components in \eqref{plane_DPDM} by
\begin{align}
    A'_{x,0}e^{-i\alpha_{x,0}}&= -\frac 1{\sqrt2} \left(\tilde A'_+-\tilde A'_- \right),\\
    A'_{y,0}e^{-i\alpha_{y,0}}&= -\frac i{\sqrt2} \left(\tilde A'_++\tilde A'_- \right),\\
    A'_{z,0}e^{-i\alpha_{z,0}}&= \tilde A'_0.
\end{align}
The presence of $f_d=1/(\textrm{sidereal day})$ in the time dependence of \eqref{DPDM_signal} arises due to the rotation of Earth.  The dark-photon amplitudes $\tilde A'_m$ are defined in the inertial celestial frame, while the signal in \eqref{DPDM_signal} is measured in the rotating Earth-fixed frame.  Translating between these frames induces a polarization-dependent shift to the frequency of the signal.  (In frequency space, this will introduce sidebands at $f=f_{A'}\pm f_d$ to the signal.)

In the case of axion DM, the effective current requires a background magnetic field $\bm B_0$.  For the Earth transducer effect, the Earth's DC magnetic field plays the role of $\bm B_0$.  We model the geomagnetic field using the IGRF-13 model~\cite{IGRF}, which provide coefficients $g_{\ell m}$ and $h_{\ell m}$ for a multipole expansion
\begin{equation}\label{eq:IGRF}
    \bm B_0=\sum_{\ell,m}C_{\ell m} \left(\frac Rr\right)^{\ell+2}\left[(\ell+1)\bm Y_{\ell m}-\bm\Psi_{\ell m}\right],
\end{equation}
where
\begin{equation}
    C_{\ell m}=(-1)^m\sqrt{\frac{4\pi(2-\delta_0^m)}{2\ell+1}}\frac{g_{\ell m}-ih_{\ell m}}2.
\end{equation}
The expansion in \eqref{IGRF} is dominated by the $C_{10}$ term, but subsequent terms lead to a $\sim10\%$ correction.  If we plug \eqref{IGRF} into \eqref{jeff_axion}, then the solution to \eqref{ampere_noE} in the spherical Earth-ionosphere cavity is
\begin{equation}\label{eq:ax_signal}
    \bm B_a(\Omega,t)=\sum_{\ell,m}\frac{(\ell+1)m_aR}{\ell(\ell+1)-(m_aR)^2}\cdot\Im\left[g_{a\gamma}\tilde a_0C_{\ell m}\cdot\bm\Phi_{\ell m}(\Omega)\cdot e^{-im_at}\right],
\end{equation}
where $\tilde a_0=a_0e^{-i\alpha_0}$ is the complex amplitude of the axion field.  Note that while the DPDM signal includes only dipole modes, the axion signal in principle receives contributions from higher modes, due to the geomagnetic field's deviation from an exact dipole.  Moreover, the axion signal does not exhibit the sidebands seen in the DPDM case.  This is because the direction of $\Jeff[,a]$ is determined by the geomagnetic field, which corotates with the Earth, whereas the direction of $\Jeff[,A']$ is determined by the DPDM polarization, which does not.

Finally, we address the case of mDM.  As with axion DM, the presence of the geomagnetic field is crucial for the Earth transducer effect.  In the mDM case, however, $\Jeff[,\phi]$ is proportional to the vector potential $\A_0$ of the Earth's field [as in \eqref{jeff_mdm}].  In temporal Coulomb gauge, the vector potential corresponding to \eqref{IGRF} takes the form
\begin{align}\label{eq:IGRF_A}
    \A_0=\sum_{\ell,m}\frac{C_{\ell m}R^{\ell+2}}{\ell r^{\ell+1}}&\Bigg(\frac{\kappa_{\ell m}R}r[(\ell+1)\bm Y_{\ell m}-\bm\Psi_{\ell m}]-\bm\Phi_{\ell m}\Bigg).
\end{align}
The final term in \eqref{IGRF_A} is fixed by observations of the geomagnetic field.  The coefficients $\kappa_\ell$, however, are not, as they lead to no observable magnetic field.  Instead, these coefficients are determined by the internal currents in the Earth that source the geomagnetic field, and so they are subject to modeling uncertainties of the Earth's interior.  \citeR{millicharged} estimated the range of the dominant dipolar coefficient to be $0.5\lesssim\kappa_{10}\lesssim2.3$.  It is clear from the form of \eqref{IGRF_A} that the terms proportional to $\kappa_{\ell m}$ will give a similar magnetic-field response as \eqref{ax_signal}.  In particular, it will depend only on $R$, not on $h$.  On the other hand, the magnetic-field response generated by the final term in \eqref{IGRF_A} will be suppressed by $h$.  This can be seen from a similar Amp\`ere's law argument as the DPDM case, shown in the right panel of \figref{ampere}, using a different Amp\`erian loop.  The dominant contribution to $\B_\phi$, therefore, comes from the model-dependent portion of \eqref{IGRF_A},
\begin{equation}\label{eq:mdm_signal}
    \bm B_\phi(\Omega,t)=\sum_{\ell,m}\frac{(\ell+1)R^2}{\ell^2(\ell+1)-\ell(m_\phi R)^2}\cdot\Re\left[e_m^2\tilde\phi_0^2\kappa_{\ell m}C_{\ell m}\cdot\bm\Phi_{\ell m}(\Omega)\cdot e^{-2im_\phi t}\right],
\end{equation}
where $\tilde\phi_0=\phi_0e^{-i\alpha_0}$ is the complex mDM amplitude.  As expected, $\B_\phi$ oscillates at frequency $2m_\phi$ (and does not exhibit sidebands from daily modulation).

Let us comment on some of the important properties of the Earth transducer signal.  As we have emphasized, it is proportional to the large radius of the Earth, $R$, as opposed to the small height of the atmosphere, $h$.  It is also spatially coherent, that is, \eqref[s]{DPDM_signal}, (\ref{eq:ax_signal}), and (\ref{eq:mdm_signal}) all have a prescribed dependence on $\Omega$, so that at a fixed time $t$, the signals at different locations on the Earth are correlated.  This is in contrast to noise, which would be uncorrelated between different locations.  Combining data from many locations therefore allows us to better discriminate our signal from noise.  In this way, we can leverage datasets with many simultaneous measurements across the Earth.  This signal is also temporally coherent, that is, it has an exact oscillatory dependence on $t$.  This allows us to also leverage datasets that span a long period of time.  As mentioned below \eqref{plane_nr}, this coherence remains valid for $t\lesssim\Tcoh$, after which the amplitude and phase of the signal will change.  Note that for the axion and mDM signals, the overall amplitude and phase will change from one coherence time to the next, but the spatial dependence will not.%
\footnote{The spatial dependences of the axion and mDM signals will, however, change as the geomagnetic field slowly drifts over time, which occurs at a rate of $\sim1\%$ per decade.}
By contrast, the spatial dependence of the DPDM signal may change, as the polarization of the DPDM can vary from coherence time to coherence time.

Finally, we discuss the robustness of the results that we have derived.  \eqref[s]{DPDM_signal}, (\ref{eq:ax_signal}), and (\ref{eq:mdm_signal}) were derived for an idealized spherical model of the Earth and ionosphere, even though, as discussed at the beginning of this subsection, the Earth's real atmospheric profile can be much more complicated.  Nevertheless, these results can be made robust to environment details.  Let us write the full magnetic-field response as
\begin{equation}
    \B_\DM=\B_\mathrm{sph}+\B_\mathrm{env},
\end{equation}
where $\B_\mathrm{sph}$ is the result derived for a spherical boundary, and $\B_\mathrm{env}$ is the additional contribution required to satsfy the true boundary conditions of the complicated near-Earth environment.  As argued earlier in \secref{ubdm}, so long as $\lambda_\DM\gg R$, the electric field in the Amp\`ere-Maxwell law can be neglected and the resulting signal $\B_\DM$ satisfies \eqref{ampere_noE}.  This implies that $\nabla\times\B_\mathrm{env}=0$.  In terms of VSH, it is clear from \eqrefRange{curlY}{curlPhi} that $\B_\mathrm{env}$ must be composed only of $\bm Y_{\ell m}$ and $\bm\Psi_{\ell m}$ contributions.  In particular, the $\bm\Phi_{\ell m}$ contribution will come solely from \eqref{DPDM_signal}, (\ref{eq:ax_signal}), or (\ref{eq:mdm_signal}) [depending on the UBDM model].  Because VSH form an orthonormal basis for vectorial functions on the sphere, then the $\bm\Phi_{\ell m}$ contribution can be separated from the $\bm Y_{\ell m}$ and $\bm\Psi_{\ell m}$ contributions through an appropriate projection.  In this way, we can isolate a part of the signal which is robust to atmospheric modeling.  The important properties of the Earth transducer signal, including how they differ between UBDM models, are summarized in \tabref{properties},

\begin{table}
    \centering
    \begin{tabular}{c@{\hskip 1.5mm}|@{\hskip 1.5mm}c@{\hskip 1.5mm}|@{\hskip 1.5mm}c@{\hskip 1.5mm}|@{\hskip 1.5mm}c}
        \hline\hline
        Property&DPDM&Axion DM&mDM\\\hline
        $\B_\DM$ scales with lengthscale&Scales with $R$&Scales with $R$&\begin{tabular}{@{}c@{}}$\bm\Phi_{\ell m}$ parts scale with $R$\\$\bm Y_{\ell m},\bm\Psi_{\ell m}$ parts scale with $h$\end{tabular}\\\hline
        Spatial dependence&\begin{tabular}{@{}c@{}}Only $\bm\Phi_{1m}$ (in NR limit)\\Fixed by DPDM polarization\end{tabular}&\begin{tabular}{@{}c@{}}All $\bm\Phi_{\ell m}$\\Fixed by multipoles of $\bm B_0$\end{tabular}&\begin{tabular}{@{}c@{}}Dominantly $\bm\Phi_{\ell m}$\\Fixed by internal currents\end{tabular}\\\hline
        Temporal dependence&\begin{tabular}{@{}c@{}}Oscillates at $f=f_{A'},f_{A'}\pm f_d$\\Phase, amplitude, \&\\[-1mm]orientation change every $\Tcoh$\end{tabular}&\begin{tabular}{@{}c@{}}Oscillates at $f=f_a$\\Phase \& amplitude\\[-1mm]change every $\Tcoh$\end{tabular}&\begin{tabular}{@{}c@{}}Oscillates at $f=2f_\phi$\\Phase \& amplitude\\[-1mm]change every $\Tcoh$\end{tabular}\\\hline
        Robustness to modeling&$\bm\Phi_{1m}$ parts robust&$\bm\Phi_{\ell m}$ parts robust&\begin{tabular}{@{}c@{}}$\bm\Phi_{\ell m}$ parts robust to atmosphere,\\[-1mm]but depend on interior\end{tabular}\\\hline\hline
    \end{tabular}
    \caption{Summary of important properties of Earth transducer signal, and how they vary for each UBDM model}
    \label{tab:properties}
\end{table}

\subsection{Higher masses}
\label{sec:high}

Let us now discuss techniques to extend the Earth transducer effect to higher DM masses.  In \secref[s]{ubdm} and \ref{sec:transducer}, we focused on the low-mass regime where $\lambda_\DM\gg R$.  This was an important condition to neglect the electric field contribution to the Amp\`ere-Maxwell law in \eqref{ampere}, and ultimately allowed us to argue that the low-mass signal was robust.  At masses $m_\DM\gtrsim3\times10^{-14}\eV$, this argument no longer holds, and environmental effects can become important.  In fact, this is known to be the case.  To see how, note that the axion DM signal in \eqref{ax_signal} exhibits resonances at masses $m_\DM=\sqrt{\ell(\ell+1)}/R$.  These are the so-called ``Schumann resonances" of the Earth-ionosphere cavity~\cite{Schumann,bliokh1980schumann,Rodriguez-Camacho}.  The idealized spherical model that we employed in \secref{transducer} predicts them at frequencies $f=10.6\Hz,18.3\Hz,25.9\Hz,\ldots$\,.  Measurements of the Schumann resonances, however, find typical values of $7.4-8.0\Hz$, $13.7-14.6\Hz$, and $19.9-21.1\Hz$ (depending on time of day and season) for the first three resonances, with widths as low as $1.5-2.0\Hz$~\cite{Rodriguez-Camacho}.  The spherical model, therefore, mispredicts the Schumann resonances by more than their widths.  As the Earth transducer effect formally diverges at these resonances, the naive spherical model could give predictions for $m_\DM\gtrsim3\times10^{-14}\eV$ which are off by orders of magnitude!  It is then clear that any computation of the global Earth transducer effect must account for atmospheric effects.

There are two approaches which have been proposed to address the high-mass regime.  \citeR{curl} proposed to perform a local measurement of the DM effective current by measuring $\nabla\times\B_\DM$ rather than $\B_\DM$.  Meanwhile, \citeR[s]{eskdalemuir_letter,eskdalemuir_theory} included atmospheric effects in the modeling of the global Earth transducer effect.  Let us begin with the first approach.  In the regime $\lambda_\DM\ll R$, the electric field response is no longer negligible compared to the magnetic-field response, and so the $\partial_t\E$ term in \eqref{ampere} cannot simply be dropped.  However, because the ground is a good conductor, then near the surface of the Earth, the component parallel to the ground, $\E_\parallel$, must still vanish!  More precisely, it is argued in \citeR{curl}, that $\E_\parallel$ is suppressed everywhere inside the Earth-ionosphere cavity, so long as $\lambda_\DM\gg h$.  Then for DM masses $3\times10^{-14}\eV\lesssim m_\DM\lesssim2\times10^{-12}\eV$, we can take the parallel component of \eqref{ampere} to find
\begin{equation}\label{eq:ampere_curl}
    (\nabla\times\B)_\parallel=\Jeff[,\parallel].
\end{equation}
\eqref{ampere_curl} gives us a reliable prediction for $\nabla\times\B_\DM$ in the directions parallel to the ground.  Importantly, this effect is insensitive to atmospheric effects, as it does not rely on solving for a response in the presence of some global boundary conditions, but rather it is a direct probe of the local DM effective current.  Note that because SM currents in the lower atmosphere are suppressed, then any environmental magnetic-field noise should satisfy $\nabla\times\B_\mathrm{noise}=0$.  Measuring $\nabla\times\bm B$, therefore, acts as a natural background rejection scheme for environmental noise.

In order to measure $\nabla\times\B_\DM$ in a horizontal direction, we require a differential measurement along a vertical gradient.  This can be achieved, for instance, near a hill or mountain. \figref{curl} shows an illustration of how such a measurement can be performed.  Three magnetic-field measurements are taken at positions $\bm r_0,\bm r_1,\bm r_2$.  The plane containing these positions must be parallel to the vertical direction.  At each position, we measure the component of the magnetic field parallel to the baseline defined by the other two positions.  These measurements can then be combined as
\begin{align}\label{eq:Delta}
    \Delta&\equiv\B(\bm r_0)\cdot(\bm r_2-\bm r_1)+\bm B(\bm r_1)\cdot(\bm r_0-\bm r_2)+\bm B(\bm r_2)\cdot(\bm r_1-\bm r_0)\\
    &\approx\left[(\bm r_2-\bm r_0)\times(\bm r_1-\bm r_0)\right]\cdot(\nabla\times\B),
\end{align}
where the approximation holds in the limit $\bm r_1,\bm r_2\rightarrow\bm r_0$.  This procedure gives a direct measurement of $\nabla\times\B$ in the direction normal to the plane containing the three positions.  Note that $\Delta$ is proportional to the area of the triangle formed by $\bm r_0,\bm r_1,\bm r_2$, so it is beneficial to maximize this area.  This procedure can be understood as a discretized line integral around the dotted line in \figref{curl}.

\begingroup
\makeatletter
\renewcommand{\fnum@figure}{FIG.~\thefigure~\emph{(from \citeR{curl})}}
\makeatother
\begin{figure*}[t]
\includegraphics[width=0.8\textwidth]{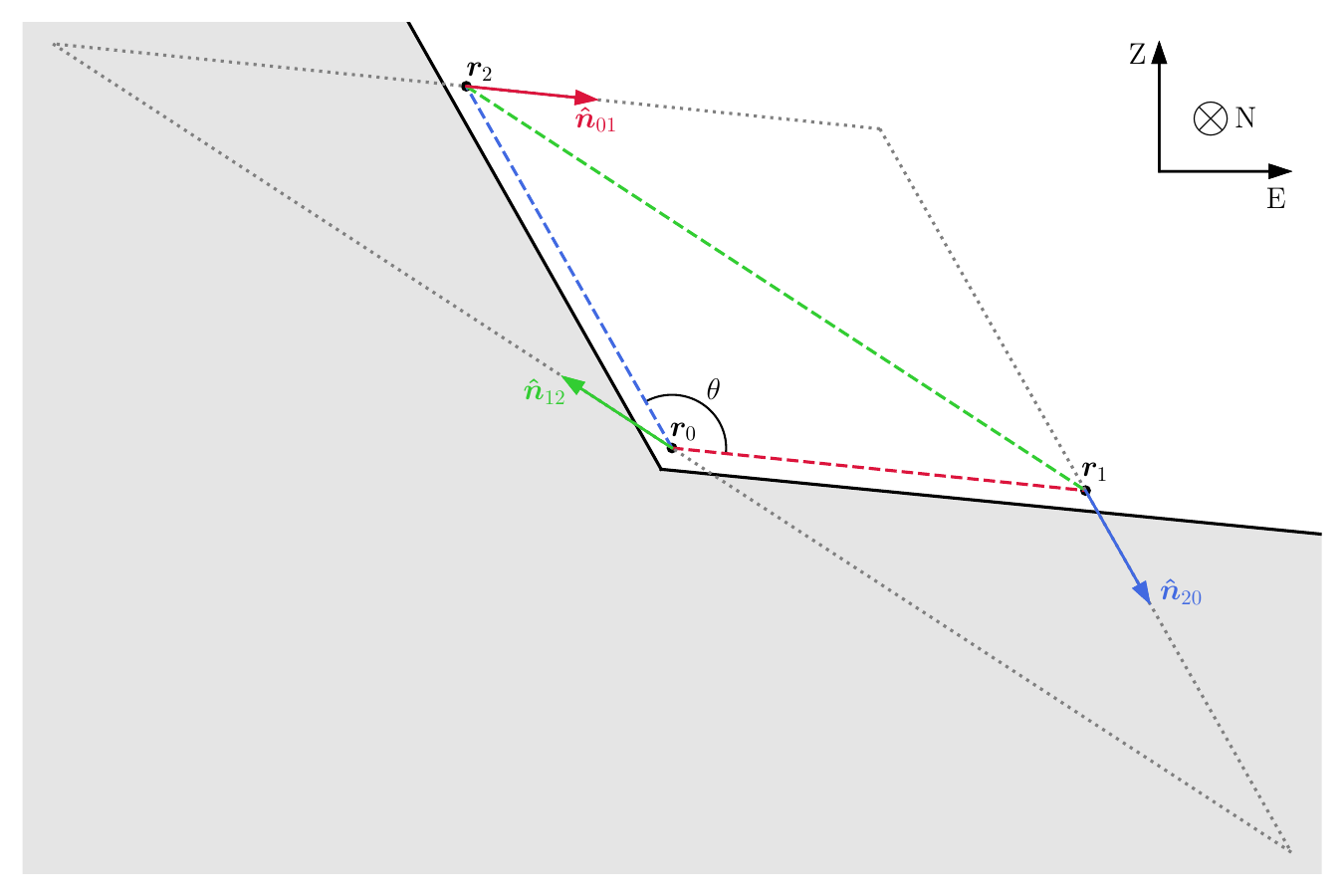}
\caption{\label{fig:curl}%
    Illustration of curl measurement scheme described in \secref{high}.  Three single-axis magnetic-field measurements are made at $\bm r_0,\bm r_1,\bm r_2$.  The direction of each measurement (represented by the unit vectors $\bm{\hat n}_{12},\bm{\hat n}_{20},\bm{\hat n}_{01}$) should be parallel to the baseline between the other two locations (shown in matching colored dashed lines).  These measurements are combined as in \eqref{Delta}, which can be understood as a discretized line integral around the dotted grey line.}
\end{figure*}
\endgroup

\citeR[s]{eskdalemuir_letter,eskdalemuir_theory,eskdalemuir_dpdm} took an alternative approach to extend the Earth transducer effect to higher masses, which was to incorporate the measured atmospheric conductivity profile into the calculation of the magnetic-field response.  This is achieved by modifying Gauss's law and the Amp\`ere-Maxwell law, \eqref[s]{gauss} and (\ref{eq:ampere}), as
\begin{align}
    \nabla\cdot(n^2\E)&=\rhoeff,\\
    \nabla\times\B-\partial_t(n^2\E)&=\Jeff,
\end{align}
where $n(\x)=\sqrt{1+i\sigma(\x)/m_a}$ is the (spatially varying) index of refraction.  \citeR[s]{eskdalemuir_letter,eskdalemuir_theory,eskdalemuir_dpdm} assumed a spherically symmetric atmospheric conductivity profile $\sigma(r)$.  While the upper layers of the Earth's atmosphere exhibit significant anisotropy, for high masses $m_\DM\gtrsim3\times10^{-4}\eV$, the magnetic-field response is suppressed above $\sim100\km$.  Therefore, the calculation is most sensitive to the properties of the lower atmsophere, which is relatively isotropic.  \citeR{eskdalemuir_theory} verified this by computing the response for multiple atmospheric models, including both day- and nighttime models, and found negligible differences in the response.  Their resulting $\B_\DM$ exhibits resonant features which agree with the observed central frequencies and widths of the Schumann resonances.

\section{Experimental searches}
\label{sec:experiment}

\begin{table}
    \centering
    \makebox[0pt][c]{\begin{tabular}{@{\hskip 1mm}c@{\hskip 1mm}|@{\hskip 1mm}c@{\hskip 1mm}|@{\hskip 1mm}c@{\hskip 1mm}|@{\hskip 1mm}c@{\hskip 1mm}|@{\hskip 1mm}c@{\hskip 1mm}|@{\hskip 1mm}c@{\hskip 1mm}}
    \hline\hline
    Property & SuperMAG & Eskdalemuir & SNIPE Hunt & Munich & GPEX\\\hline
    Type & Existing dataset & Existing dataset & Dedicated search & Dedicated search & Dedicated search\\
    Total number of sites & $\sim500$ & 1 & 5 & 1 & 1\\
    Total duration & $\sim50$\,yrs & $\sim10$\,yrs & $\sim80$\,hrs & $\sim10$\,hrs & $\sim3$\,hrs\\
    Frequency range [Hz] & $6\times10^{-4}$--$1$ & $0.02$--$40$ & $0.5$--$5$ & $5\times10^{-3}$--$7.5$%
    \footnote{We quote (and show in \figref{constraints}) the frequency range stated in \citeR{munich}, but recall that the Earth transducer signal calculation becomes unreliable near the first Schumann resonance.  Constraints in the $5$--$7.5$\,Hz range should therefore be viewed with uncertainty.}
    & $0.1$--$5$\\
    Mass range [eV] & $2\times10^{-18}$--$4\times10^{-15}$ & $7\times10^{-17}$--$2\times10^{-13}$ & $2\times10^{-15}$--$2\times10^{-14}$ & $2\times10^{-17}$--$3\times10^{-14}$ & $4\times10^{-16}$--$2\times10^{-14}$\\\hline\hline
    \end{tabular}}
    \caption{Comparison of various experimental searches for the Earth transducer effect}
    \label{tab:searches}
\end{table}

We have seen that the Earth transducer effect generates a global AC magnetic-field signal at the Earth's surface.  The simplest way to detect this effect is with data from unshielded magnetometers.  In order to leverage the spatial and temporal coherence discussed in \secref{transducer}, it is beneficial for this dataset to include measurements from many magnetometers over a long period of time.  In this section, we review several searches for the Earth transducer effect in such datasets.  A particularly powerful strategy has been to repurpose existing datasets to search for UBDM through this effect.  \citeR[s]{dpdm_search,axion,supermag_1sec} searched a publicly available dataset maintained by the SuperMAG collaboration~\cite{SuperMAGwebsite,Gjerloev:2009wsd,Gjerloev:2012sdg}, which collates magnetic-field data from a global array of unshielded magnetometers.  Likewise, \citeR[s]{eskdalemuir_letter,eskdalemuir_analysis,eskdalemuir_dpdm} searched magnetic-field data from the Eskdalemuir observatory~\cite{eskdalemuir} for the Earth transducer effect.  At frequencies $f_\DM\gtrsim\Hz$ ($m_\DM\gtrsim4\times10^{-15}\eV$), much of the ambient magnetic-field noise is anthropogenic~\cite{Constable}, so noise levels can be reduced significantly by performing measurements in a radio quiet location.  For this reason, the SNIPE Hunt collaboration~\cite{snipe_hunt,snipe_hunt_2026}, Munich group~\cite{munich}, and GPEX collaboration~\cite{gpex} have performed dedicated high-frequency searches for the Earth transducer effect in remote locations.  The properties of these searches are compared in \tabref{searches}.  All of these searches have placed constraints on UBDM parameter space, as summarized in \figref[s]{constraints} and \ref{fig:constraints_mdm}.  In many regions of parameter space, unshielded magnetometers are the most sensitive experimental probes of UBDM.

\subsection{Existing datasets}
\label{sec:existing}

Low-frequency ground-based magnetometers are frequently employed to study geophysical processes and solar activity.  As such, an abundance of unshielded magnetometer data has already been recorded.  One prominent example is the publicly available data maintained by the SuperMAG collaboration~\cite{SuperMAGwebsite,Gjerloev:2009wsd,Gjerloev:2012sdg}.  This collaboration aggregates and processes data from an array of magnetometer stations dispersed around the globe.  They maintain two datasets: a ``low-fidelity" (1-minute time resolution) dataset, which combines data from $\sim500$ stations over a span of more than 50 years; and a ``high-fidelity" (1-second time resolution) dataset, which combines data from $\sim200$ stations over more than 20 years.  \citeR[s]{dpdm_search,axion} performed searches of the SuperMAG low-fidelity dataset for the Earth transducer signal (for DPDM and axion DM, respectively) at Compton frequencies $f_\DM<1/(1\,\mathrm{min})$, corresponding to masses $m_\DM\lesssim7\times10^{-17}\eV$, while \citeR{supermag_1sec} performed a search of the high-fidelity dataset (for both DPDM and axion DM) for $f_\DM<1\Hz$, corresponding to $m_\DM\lesssim4\times10^{-15}\eV$.

Here we give a cursory overview of the analysis framework utilized in these searches.  The ideal search protocol would be to Fourier transform the magnetic-field timeseries data from each station, combine them to project onto the VSH combination in \eqref{DPDM_signal} or (\ref{eq:ax_signal}), and then search for peaks in the resulting spectrum.  The SuperMAG datasets are, however, complicated by the fact that most stations do not actively report data for the entire duration of the dataset.  These periods of inactivity make a naive Fourier transform less straightforward.  Moreover, the noise levels vary significantly from station to station, so that combining data from different stations with comparable weights may be suboptimal.  Instead, the strategy employed by \citeR[s]{dpdm_search,axion,supermag_1sec} was as follows.  First, each station's measurements are weighted by a VSH motivated by the Earth transducer signal, e.g., schematically
\begin{align}
    X_i^{(n)}(t_j)\sim\Phi_{\ell m}^\alpha(\Omega_i)\cdot B_i^\alpha(t_j),
\end{align}
where $B_i^\alpha(t_j)$ represents the $\alpha$ component of the magnetic-field data from station $i$ at time $t_j$, $\Phi_{\ell m}^\alpha$ represents the $\alpha$ component of $\bm\Phi_{\ell m}$, and $\Omega_i$ represents the location of station $i$.  The $n$ quantities $X_i^{(n)}$ weight the data from station $i$ with different VSH components and coefficients.  The quantities from each station are then combined into $n$ timeseries, weighted by the noise levels in each station, e.g., schematically
\begin{equation}\label{eq:Xn}
    X^{(n)}(t_j)\sim\frac1{W(t_j)}\sum_iw_iX_i^{(n)}(t_j),
\end{equation}
where $w_i$ are station weights that are chosen inversely proportional to the station's DC noise level, and the sum runs over all $i$ which report a measurement at time $t_j$.  The total weights $W(t_j)=\sum_iw_i$ can depend on $t_j$ because the number of active stations vary with time.  In this way, the framework constructs a small number $n$ of timeseries with nonzero values at all $t$.  The combination procedure weights station data both according to their noise level and the prediction in \eqref{DPDM_signal} or (\ref{eq:ax_signal}).%
\footnote{As mentioned in \secref{transducer}, for low masses $m_\DM\ll10^{-16}\eV$, the complicated geometry of Earth's atmosphere may introduce $\bm Y_{\ell m}$ and $\bm\Psi_{\ell m}$ contributions into \eqref{DPDM_signal} or (\ref{eq:ax_signal}).  In principle, these can be removed via a projection onto the $\bm\Phi_{\ell m}$ VSH.  However, because the combination procedure in \eqref{Xn} includes the weights $w_i$, and because the stations are not uniformly distributed across the Earth, the sum in \eqref{Xn} does not constitute an exact VSH projection.  In \citeR{dpdm_search}, it was estimated that this effect may alter the signal prediction, as well as the resulting constraint, by an $\O(1)$ factor for the low-fidelity search.}
These $n$ timeseries then become the primary object of study for the search procedure, namely, they are Fourier transformed and compared to noise in a Bayesian framework to determine the significance of a potential signal.  This framework requires both a prediction for how the signal would appear in $X^{(n)}$ and an estimate for the noise in $X^{(n)}$.  To determine the signal prediction, the same combination procedure that is performed on the data $B_i^\alpha(t_j)$ is applied to the predicted magnetic-field signal in \eqref{DPDM_signal} or (\ref{eq:ax_signal}).  The noise, on the other hand, is determined by a data-driven analysis directly of the timeseries $X^{(n)}$.

The analyses in \citeR[s]{dpdm_search,axion,supermag_1sec} each found a number of signal candidates which exhibited high statistical significance.  Many such candidates were the result of transient anomalies in single stations (either from the instrument itself or from post-processing of the data).  By contrast, a true UBDM signal would be present in all stations at all times.  \citeR[s]{dpdm_search,axion,supermag_1sec} applied secondary checks, such as geographic uniformity and temporal persistence, to each of these candidates and ultimately concluded that none consituted robust evidence for UBDM.  As a result, they placed constraints on DPDM and axion DM parameter space.  The green and blue curves in \figref{constraints} show the smoothed constraints from the low and high-fidelity searches, respectively.

Another series of works has analyzed data from the Eskdalemuir observatory in Scotland~\cite{eskdalemuir}.  \citeR[s]{eskdalemuir_letter,eskdalemuir_analysis} searched this dataset for an axion DM signal, while \citeR{eskdalemuir_dpdm} considered a DPDM signal.  Importantly, both of these searches included atmospheric modeling into their signal computation which allowed for a robust prediction of the signal up to $f_\DM\sim30\Hz$.  In the absence of any robust signal candidates, they also set constraints on DPDM and axion DM parameter space, shown in purple in \figref{constraints}.%
\footnote{The analyses in \citeR[s]{eskdalemuir_analysis,eskdalemuir_dpdm} discard narrow frequency ranges around integer frequencies and very narrow ranges around multiples of $0.05\Hz$.  As a result, their constraints do not rule out all frequencies in the regions shown in \figref{constraints}.  Rather these constraints should be understood to have many small gaps.}
Notice that the sensitivity peaks near the first Schumann resonance $f_\DM\sim7\Hz$, as this is where the Earth transducer signal is maximized.

\subsection{Dedicated searches}
\label{sec:dedicated}

\begin{table}
    \centering
    \begin{tabular}{@{\hskip 3mm}c@{\hskip 3mm}|@{\hskip 3mm}c@{\hskip 3mm}|@{\hskip 3mm}c@{\hskip 3mm}}
    \hline\hline
    Property & SNIPE Run I & SNIPE Run II \\\hline
    Number of sites & 3 & 5\\
    Total (analyzed) duration & $\sim60$ ($\sim25$) hrs & $\sim80$ ($\sim15$) hrs \\
    Sensor type & Giant magnetoresistance & Induction coil \\
    Sensor bandwidth &  DC to $80\Hz$ & $<0.1\Hz$ to $1.0\Hz[k]$\\
    Noise floor & $\sim300\,\mathrm{pT/\sqrt{Hz}}$ & $\sim1\,\mathrm{pT/\sqrt{Hz}}$\\
    Detection axes& 3 (N-S, E-W, Vertical) & 2 (N-S, E-W)\\
    Data acquisition system & Laptops with serial connection to Twinleaf VMR & Raspberry Pi + Pi-Plates ADC board\\\hline\hline
    \end{tabular}
    \caption{Characteristics of SNIPE Runs I and II}
    \label{tab:snipe}
\end{table}

At higher frequencies $f_\DM\gtrsim\Hz$, ambient magnetic field noise is dominated by anthropogenic sources~\cite{Constable}.  This implies that the choice of location is increasingly important for high-frequency measurements of the Earth transducer effect.  The SNIPE Hunt collaboration was formed to take dedicated measurements which could fully leverage this observation.  SNIPE Hunt has undertaken campaigns in 2022 (SNIPE Run I)~\cite{snipe_hunt} and 2025 (SNIPE Run II)~\cite{snipe_hunt_2026,snipe_hunt_mdm}.  The goal of each campaign was to acquire time-synchronized measurements of magnetic fields in regions with minimal man-made magnetic interference, such as arises from power lines and moving vehicles.  \tabref{snipe} contrasts the relevant characteristics of these two runs.  In each case, magnetic field measurements were simultaneously performed at a number of geographically separated sites.%
\footnote{In contrast to the SuperMAG searches, the SNIPE Hunt runs only took measurements at a few sites, and the Munich and GPEX searches only had one site each.  This means these searches cannot reliably perform the VSH projection mentioned at the end of \secref{transducer}.  Because these searches focus on masses $m_\DM\gg10^{-16}\eV$, we expect the spherical Earth-ionosphere model to remain valid for their mass range of interest.  Therefore, \eqref[s]{DPDM_signal}, (\ref{eq:ax_signal}), and (\ref{eq:mdm_signal}) can be used without the VSH projection.}
The acquired data was time-stamped with GPS measurements and written to file for subsequent analysis.  The sensors and data acquisition system were powered by rechargeable lithium-ion battery packs, which were periodically replaced during the science runs as they discharged.

In SNIPE Run I, each site deployed Twinleaf Vector Magnetoresistive (VMR) magnetometers~\cite{twinleaf}.  The noise floor of the measurement was limited by the instrumental noise, which in this case, was frequency independent.  The analysis of the SNIPE Run I data followed a similar Bayesian framework to the one employed in the SuperMAG searches.  Data was modeled as a signal, given by \eqref{DPDM_signal} or (\ref{eq:ax_signal}), on top of Gaussian noise.  In the Run I analysis, a frequency-independent data-driven estimate was used for the noise.

In SNIPE Run II, two of the five total sites deployed induction-coil magnetometers supplied by Lemi Sensors LLC~\cite{lemi}, while the other three sites developed their own sensors and readout electronics in their labs.  At the heart of each sensor was $\O(10^4)$ turns of copper wound around a mu-metal core.  A flux-feedback approach was used to flatten the response across the desired bandwidth.  The noise floor for the sensors used in SNIPE Run II was set by a combination of instrumental and environmental noise.  Consequently, this noise displayed non-negligible frequency dependence and non-stationarity.  In the analysis, a frequency- and time-dependent noise model was applied based on a running average of power in different frequency bins.

\begin{figure*}[t]
\includegraphics[width=0.49\textwidth]{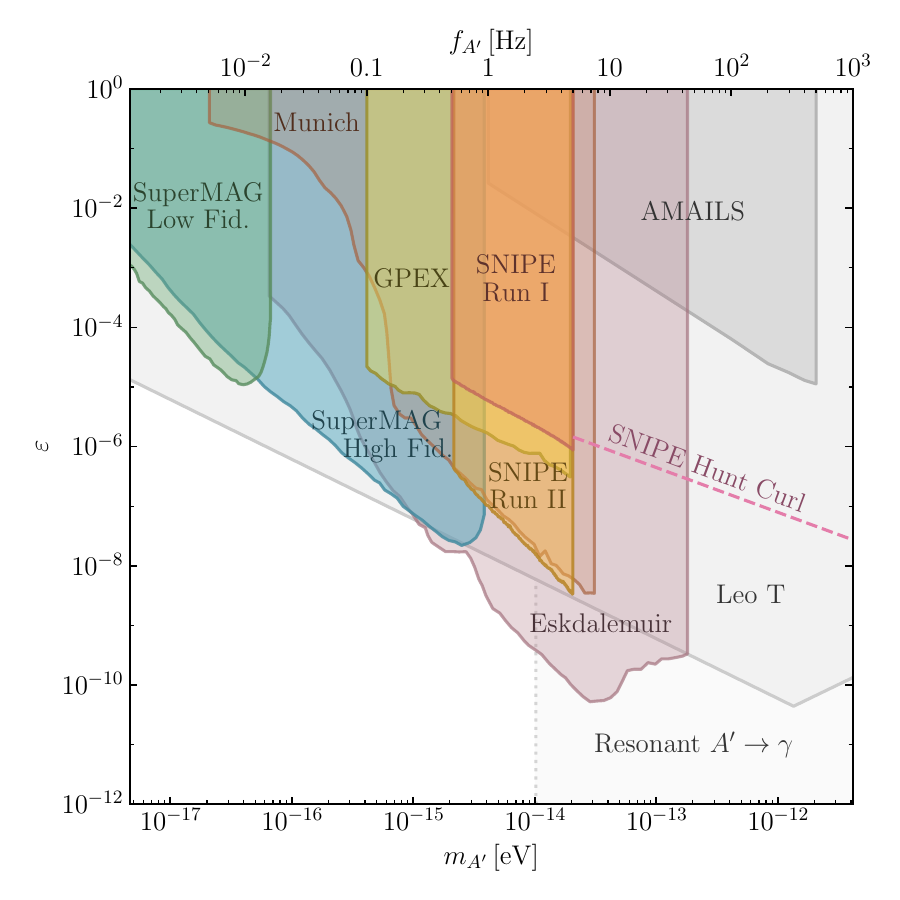}
\includegraphics[width=0.49\textwidth]{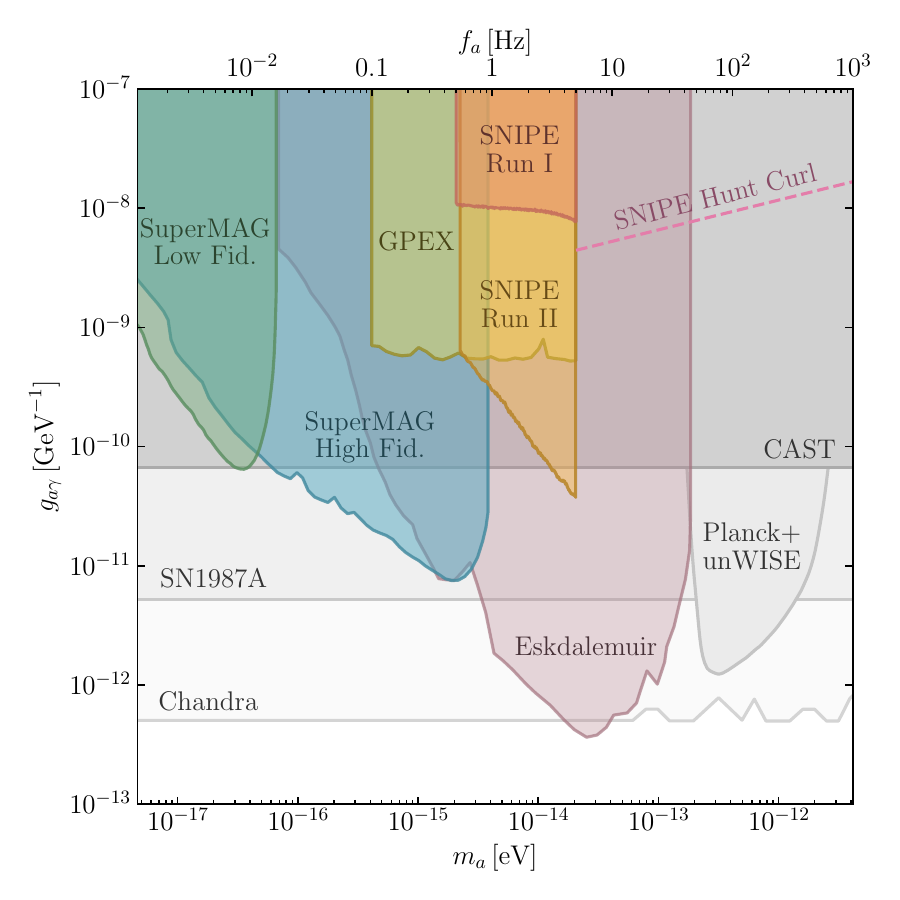}
\caption{\label{fig:constraints}%
    Summary of constraints on DPDM (\emph{left}) and axion DM (\emph{right}) from searches for the Earth transducer effect.  In cool colors, we show smoothed constraints from searches in existing datasets: the SuperMAG low- and high-fidelity datasets~\cite{dpdm_search,axion,supermag_1sec,SuperMAGwebsite,Gjerloev:2009wsd,Gjerloev:2012sdg}, and data from the Eskdalemuir observatory~\cite{eskdalemuir_analysis,eskdalemuir_dpdm,eskdalemuir}.  In warm colors, we show smoothed constraints from dedicated searches by various collaborations: two data-taking runs by the SNIPE Hunt collaboration~\cite{snipe_hunt,snipe_hunt_2026}, one run by the Munich group~\cite{munich}, and one run by the GPEX collaboration~\cite{gpex}.  The dashed pink lines show the sensitivities that SNIPE Hunt could achieve with the curl measurement discussed in \secref{high}~\cite{curl}.  In grey, we show various existing constraints, with darker shades indicating direct constraints and lighter shades indicating astrophysical/cosmological constraints (see text for details).}
\end{figure*}

Both SNIPE Runs were able to set constraints on DPDM and axion DM in the frequency range $0.5\Hz\lesssim f_\DM\lesssim5\Hz$, with the upper bound chosen to avoid atmospheric dependence from the Schumann resonances.  These constraints are shown in red (Run I) and orange (Run II) in \figref{constraints}.  Other groups have performed similar dedicated searches for the Earth transducer effect.  A group from the Technical University of Munich~\cite{munich} conducted a search for DPDM in the Austrian Alps for a duration of 10.5 hours using Twinleaf Pulsed Pump Magnetometers (PPM)~\cite{twinleaf_ppm}.  In addition, the GPEX collaboration~\cite{gpex} performed a search for axion DM and DPDM in the desert of XiaoDushan in Gansu Province, China for a duration of one hour using an atomic magnetometer manufactured by QUSPIN~\cite{quspin}.  The results from their searches are shown in brown and yellow, respectively, in \figref{constraints}.  In the future, the SNIPE Hunt collaboration intends to implement the curl technique described in \secref{high} and illustrated in \figref{curl}.  The dashed pink lines in \figref{constraints} show the projections for SNIPE Hunt's implementation of the curl measurement, based on \citeR{curl}.

\begin{figure*}[t]
\includegraphics[width=0.49\textwidth]{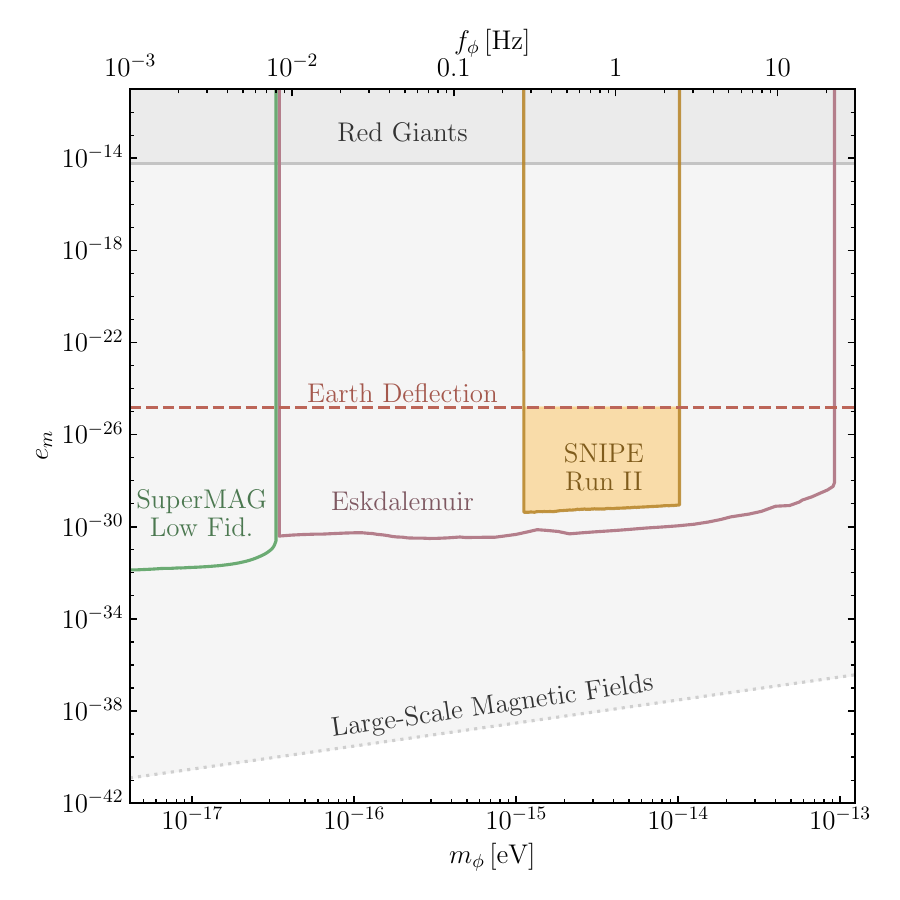}
\caption{\label{fig:constraints_mdm}%
    Sensitivity of Earth transducer effect to mDM.  We recast the SuperMAG low-fidelity~\cite{axion} and Eskdalemuir~\cite{eskdalemuir_analysis} axion DM constraints from \figref{constraints} to mDM parameter space via the mapping in \eqref{recast} [with $\kappa_{\ell m}=\kappa_{10}=0.5$].  These senstivity curves are not full reanalyses of the corresponding datasets and depend on modeling of the Earth's interior.  Nevertheless, they can be interpetted, up to an $\O(1)$ factor, as constraints on mDM parameter space.  In orange, we show a dedicated analysis of the SNIPE Run II dataset for an mDM signal~\cite{snipe_hunt_mdm}.  We shade this region to indicate a more rigorous exclusion.  Above the dashed red line, the Earth's magnetic field can deflect mDM significantly (see footnote \ref{ftnt:deflection}), so that the treatment in this review no longer holds.  In grey, we also show existing astrophysical/cosmological constraints (see text for details).  Note that the constraint from Large-Scale Magnetic Fields is significantly relaxed if mDM is a subcomponent of DM~\cite{Kadota_2016,Stebbins_2019}.}
\end{figure*}

\section{Discussion and conclusion}
\label{sec:conclusion}

In this review, we have described how the Earth can act as a transducer for EM-coupled UBDM, allowing for its potential detection by unshielded magnetometers.  The effect of UBDM can be parametrized as an effective current $\Jeff$, as in \eqref[s]{jeff_DPDM}, (\ref{eq:jeff_axion}), and (\ref{eq:jeff_mdm}) [for various models].  These currents source AC magnetic fields via Amp\`ere's law in \eqref{ampere_noE}.  The resulting magnetic-field signal is enhanced by the large radius of the Earth, $R$, and it is spatially and temporally coherent.  We also showed that, for $m_\DM\lesssim1/R$, the signal prediction can be made robust to environmental details via a VSH projection.  We discussed various methods to extend this effect to higher frequencies, such as differential measurements of $(\nabla\times\B)_\parallel$, or inclusion of atmophseric modeling.

\figref{constraints} summarizes the results of various searches for the Earth transducer effect.  Regions shaded in cool colors represent smoothed constraints from searches in existing datasets (SuperMAG Low Fid., SuperMAG High Fid., and Eskdalemuir), while regions shaded in warm colors represent dedicated searches for the Earth transducer effect (SNIPE Run I, SNIPE Run II, Munich, and GPEX).  The dashed pink lines show the potential sensitivity of the curl measurement described in \secref{high}.  We compare these to existing constraints in various shades of grey, with darker shades representing direct constraints (AMAILS~\cite{Jiang_2024} and CAST~\cite{Anastassopoulos:2017ftl}), and lighter shades representing astrophysical/cosmological constraints (Leo T~\cite{Wadekar_2021}, Resonant $A'\rightarrow\gamma$~\cite{McDermott_2020},%
\footnote{This constraint relies on resonant conversion of DPDM into photons during the dark ages.  This mechanism was recently called into question in \citeR{Hook_2025} because nonlinearities may disrupt the resonant conversion.  For completeness, we still include this constraint in \figref{constraints}, but represent it with a dotted line.}
SN1987A~\cite{Hoof_2023}, Planck + unWISE~\cite{goldstein_2025}, and Chandra~\cite{chandra}).  The Earth transducer effect is the most powerful direct probe of DPDM with masses $m_{A'}\lesssim10^{-13}\eV$.  The constraint from the Eskdalemuir search even surpasses all astrophysical/cosmological constraints of DPDM in some mass ranges.  For axion DM, several searches for the Earth transducer effect have surpassed the leading direct constraint from CAST, while the Eskdalemuir search has probed previously unconstrained axion DM paramter space.

In \citeR{millicharged}, it was pointed out that all of these searches are also sensitive to mDM.  A comparison of \eqref[s]{ax_signal} and (\ref{eq:mdm_signal}) shows that the axion and mDM signals are approximately related by the mapping
\begin{equation}\label{eq:recast}
    e_m^2\sim\frac{\ell m_\phi^2}{\kappa_{\ell m}\sqrt{2\rho_\DM}R}\cdot g_{a\gamma}
\end{equation}
(with $m_\phi=m_a/2)$.  A precise search for the mDM Earth transducer effect, in principle, differs from a search for the axion DM effect because the linear combinations of VSH in \eqref[s]{ax_signal} and (\ref{eq:mdm_signal}) may not be the same.  However, because both expansions are dominated by the $\ell=1,m=0$ term, it is sufficient to understand the potential sensitivity to mDM by recasting the axion constraints in \figref{constraints} using \eqref{recast} with a fiducial value for $\kappa_{10}$.  In \figref{constraints_mdm}, we show recasted constraints for a couple of these searches using $\kappa_{10}=0.5$.  While these are not rigorous reanalyses of the datasets, the constraints should nevertheless be accurate to $\O(1)$.  The SNIPE Hunt collaboration has performed a dedicated analysis of their SNIPE Run II dataset for an mDM signal~\cite{snipe_hunt_mdm} (considering only the $\ell=1,m=0$ term).  We show this as a shaded orange region in \figref{constraints_mdm} to demonstrate a more rigorous exclusion.  The dashed red line in \figref{constraints_mdm} shows where the Earth's magnetic field can significantly alter the mDM distribution (see footnote \ref{ftnt:deflection}).  The treatment in this review remains valid below this line.  In grey, we also show existing astrophysical constraints from Red Giants~\cite{Davidson_2000} and Large-Scale Magnetic Fields~\cite{Kadota_2016,Stebbins_2019}.  Note that the latter constraint is significantly relaxed if the mDM is a subcomponent of DM.  In this case, the Earth transducer effect (which is relaxed to a lesser degree) becomes the strongest probe of mDM.

\acknowledgments

IFAE is partially funded by the CERCA program of the Generalitat de Catalunya.  S.K. is supported by ERC grant ERC-2024-SYG 101167211 and is funded by the European Union.  Views and opinions expressed are however those of the author only and do not necessarily reflect those of the European Union or the European Research Council Executive Agency.  Neither the European Union nor the granting authority can be held responsible for them.

I.A.S. acknowledges support from the U.S. National Science Foundation under grant \#2510628.

We would like to dedicate this review to Derek F.~Jackson Kimball and Jason E.~Stalnaker, in honor of their $50^\mathrm{th}$ birthdays and in appreciation for their invaluable contributions to this experimental program.  Derek proposed the original motivation for this research direction and ultimately coined the name ``Earth as a transducer".  Both Derek and Jason were cofounders of the SNIPE Hunt collaboration, and because of their early pioneering efforts, this experimental technique has garnered broader interest and participation. They have led and helped build a research effort that has introduced over a dozen students to aspects of experimental physics ranging from hardware development, data analysis, and field work involving hiking and camping in order to collect data in search of elusive evidence of new physics. Here's to more years of building, searching, and learning.

\appendix

\section{Vector spherical harmonics}
\label{app:VSH}

This appendix, which defines the VSH conventions used in this work, is reproduced from \citeR{dpdm_theory} with minor modifications for the convenience of the reader.

The vector spherical harmonics are defined in terms of the scalar spherical harmonic $Y_{\ell m}$ by the relations
\begin{align}
\bm{Y}_{\ell m} &= Y_{\ell m}\rhat, &
\bm{\Psi}_{\ell m} &= r\bm{\nabla} Y_{\ell m}, &
\bm{\Phi}_{\ell m} &= \bm{r}\times\bm{\nabla} Y_{\ell m},
\end{align}
where $\rhat$ is the unit vector in the direction of $\bm{r}$.  Thus $\bm{Y}_{\ell m}$ points radially, while $\bm{\Psi}_{\ell m}$ and $\bm{\Phi}_{\ell m}$ point tangentially.  Some of their relevant properties (and our phase conventions) are
\begin{align}
\bm{Y}_{\ell,-m}&=(-1)^m\bm{Y}_{\ell m}^*,\\
\bm{\Psi}_{\ell,-m}&=(-1)^m\bm{\Psi}_{\ell m}^*,\\
\bm{\Phi}_{\ell,-m}&=(-1)^m\bm{\Phi}_{\ell m}^*, \label{eq:phaseConventionPointer}\\
\bm{Y}_{\ell m}\cdot\bm{\Psi}_{\ell m}&=\bm{Y}_{\ell m}\cdot\bm{\Phi}_{\ell m}=\bm{\Psi}_{\ell m}\cdot\bm{\Phi}_{\ell m}=0,
\end{align}
\begin{align}
\int d\Omega\,\bm{Y}_{\ell m}\cdot \bm{Y}_{\ell'm'}^*&=\delta_{\ell\ell'}\delta_{mm'},\\
\int d\Omega\,\bm{\Psi}_{\ell m}\cdot\bm{\Psi}_{\ell'm'}^*&=\int d\Omega~\Phi_{\ell m}\cdot\Phi_{\ell'm'}^*\nonumber \\
	&=\ell(\ell+1)\delta_{\ell\ell'}\delta_{mm'},\\
\int d\Omega\,\bm{Y}_{\ell m}\cdot\bm{\Psi}_{\ell'm'}^*&=\int d\Omega\,\bm{Y}_{\ell m}\cdot\bm{\Phi}_{\ell'm'}^*\nonumber \\
	&=\int d\Omega\,\bm{\Psi}_{\ell m}\cdot\bm{\Phi}_{\ell'm'}^*=0.\label{eq:orthogonality}
\end{align}

For any radially dependent function $f(r)$, the divergences and curls of the VSH are given by
\begin{align}
    \nabla \cdot \left(f\bm{Y}_{\ell m}\right)&=\left( \frac{df}{dr} + \frac{2f}{r} \right) Y_{\ell m},\\
    \nabla \cdot \left(f\bm{\Psi}_{\ell m}\right)&= -\ell(\ell+1) \frac{f}{r} Y_{\ell m},\\
    \nabla \cdot \left(f\bm{\Phi}_{\ell m}\right)&=0,\label{eq:Phildiv}\\[3ex]
    \nabla\times\left(f\bm{Y}_{\ell m}\right)&=-\frac fr\bm{\Phi}_{\ell m},\label{eq:curlY}\\
    \nabla\times\left(f\bm{\Psi}_{\ell m}\right)&=\left(\frac{df}{dr}+\frac fr\right)\bm{\Phi}_{\ell m},\\
    \nabla\times\left(f\bm{\Phi}_{\ell m}\right)&=-\frac{\ell(\ell+1)f}r\bm{Y}_{\ell m}-\left(\frac{df}{dr}+\frac fr\right)\bm{\Psi}_{\ell m},\label{eq:curlPhi}
\end{align}
with the Laplacians then being
\begin{align}
    \nabla^2\left(f\bm{Y}_{\ell m}\right)
        &=\left(\frac1{r^2}\frac d{dr}\left(r^2\frac{df}{dr}\right)-\frac{(\ell(\ell+1)+2)f}{r^2}\right)\bm{Y}_{\ell m}
        \nl\quad +\frac{2f}{r^2}\bm{\Psi}_{\ell m},\\
    \nabla^2\left(f\bm{\Psi}_{\ell m}\right)
        &=\left(\frac1{r^2}\frac d{dr}\left(r^2\frac{df}{dr}\right)-\frac{\ell(\ell+1)f}{r^2}\right)\bm{\Psi}_{\ell m}
        \nl\quad +\frac{2\ell(\ell+1)f}{r^2}\bm{Y}_{\ell m},\\
    \nabla^2\left(f\bm{\Phi}_{\ell m}\right)
        &=\left(\frac1{r^2}\frac d{dr}\left(r^2\frac{df}{dr}\right)-\frac{\ell(\ell+1)f}{r^2}\right)\bm{\Phi}_{\ell m}.
\end{align}

The explicit expressions for the spherical harmonics that are relevant to the DPDM signal [see \eqref{DPDM_signal}] are
\begin{align}
\bm{\Phi}_{1,-1}(\bm{r}) &=\sqrt{\frac3{8\pi}}e^{-i\varphi}(i\thetahat+\cos\theta\phihat), \label{eq:Phi1m1}\\
\bm{\Phi}_{10}(\bm{r})&=-\sqrt{\frac3{4\pi}}\sin\theta\phihat,\label{eq:Phi10}\\
\bm{\Phi}_{11}(\bm{r})&=\sqrt{\frac3{8\pi}}e^{i\varphi}(i\thetahat-\cos\theta\phihat)\label{eq:Phi1p1},
\end{align}
where $\thetahat$ and $\phihat$ are unit vectors in the directions of increasing $\theta$ and $\varphi$.  (See footnote~\ref{ftnt:coords} for coordinate conventions.)

\bibliographystyle{JHEP}
\bibliography{references.bib}

\end{document}